\documentstyle[11pt,a4]{article}
\newcommand{\beq}{\begin{equation}}
\newcommand{\eeq}{\end{equation}}
\newcommand{\bear}{\begin{eqnarray}}
\newcommand{\earn}{\end{eqnarray}}
\author{L.I. Tsaregorodtsev and N.N. Medvedev}
\title{Spectrum  of Radiation of a Classical Electron Moving in
    the de Sitter Spacetime}

\bigskip

\begin{document}
\maketitle
\begin{abstract}
The radiation spectrum of a classical charged particle (electron)
moving in the de Sitter universe, has been calculated.
The de Sitter metric is taken in the quasi-Euclidean Robertson-Walker form.
It is shown that in the de Sitter spacetime an electron radiates
as if it moved in a constant and homogeneous electric field
with the strength vector {\bf E} collinear to the electron momentum~{\bf p}.
\end{abstract}


\section{Introduction}  

In the studies of quantum effects of field interaction in
curved spacetimes, the problem of the choice of the initial and final vacuum
states of the quantum fields is known to arise.
The choice of a vacuum to a considerable extent depends on the choice of the
frame of reference since this choice leads to a natural
criterion for the choice of a quantum field vacuum.

For the de Sitter spacetime both static and nonstatic frames of reference
can be constructed.  There are nonstatic frames where the line element
for the de Sitter space $d{\tilde{s}}^2$ has the Robertson-Walker form where
spatial sections of the de Sitter manifold can represent open hyperboloids,
three-spheres or three-planes. In the latter case the metric is called
quasi-Euclidean.  A review of properties of coordinate systems for the
de Sitter spacetime can be found in \cite{1}.

In the calculations of quantum effects of field interaction in the
Robertson-Walker spacetimes the quasi-Euclidean form of the metric is often
used.  This choice is explained by the opportunity to apply
the well developed $S$-matrix formalism of quantum
electrodynamics in an external gravitational field \cite{2'}. Besides, the
results of the calculations can be interpreted in terms of particles
moving in flat spacetime in the presence of an external gravitational
field.

The quasi-Euclidean, explicitly conformally flat form of the metric the
de Sitter spacetime is
\bear
    d{\tilde{s}}^2 = a^2(\eta)ds^2,\nonumber\\
    ds^2 = {(d\eta)}^2-{(dx^1)}^2-{(dx^2)}^2-{(dx^3)}^2,\\  
    -\infty<\eta<\infty.\nonumber
\earn
Here $ds^{2}$ is the Minkowski metric and $a(\eta)$ is the expansion law
of the Universe,
\beq
    a(\eta)=\frac{\alpha}{\eta},\ \ \ \ \  \alpha=const.
\eeq

A natural criterion for the choice of a vacuum state of a quantum massive
field in the de Sitter space with the metric (1)-(2) is the requirement for
the vacuum to be adiabatic \cite{1}. However, in the case under
consideration the metric has a coordinate singularity at $\eta=0$ and the
requirement for the vacuum to be adiabatic does not determine the choice of
the initial and final vacua in a unique manner. The ambiguity of
vacuum choice in the de Sitter space with the metric (1)-(2) makes it
expedient to study the effects of interaction of a charged particle with an
electromagnetic field in the framework of classical
electrodynamics in curved spacetime.  The results obtained can serve as an
additional criterion for the vacuum choice in quantum theory. Namely,
when the curvature of the de Sitter spacetime is small and the particle
momenta are large, the results of calculations of quantum processes having
classical analogues should coincide with the results obtained
in the framework of classical theory.

In this paper we consider the bremsstrahlung pro\-cess for a classical
charged particle moving in the de  Sitter universe with the metric (1)-(2).
The plan of the paper is as follows. In Sec. 2 we consider the motion law
of a free charged particle in the de Sitter spacetime with the metric
(1)-(2). Sec. 3 is devoted to a search for the field created by this
particle at large distances from it.  In Sec. 4 we calculate the spectral
distribution of the energy radiated from the charged particle
and analyze the dependence of the spectrum
on the spacetime curvature. Finally, in Sec. 5 we discuss our results.

\section{The particle motion law}

Let us consider the Lagrange function of a particle moving in
the curved spacetime (1)--(2):
$$
    L=-m\frac{d\tilde{s}}{d\eta}=-ma(\eta)\sqrt{1-
            \biggl(\frac{d{\bf r}}{d\eta}\biggr)^2}.
$$
We would like to use in the manifold to be examined,  instead of
the metric $d\tilde{s}^{2}$, the conformally related
metric of the Minkowski spacetime $ds^{2}$.
Then  $\bf{v}=d{\bf r}/d\eta=\dot{r}$  is
the particle velocity in an external
gravitational field relative to the flat metric. Differentiating the
Lagrange function with respect to $ {\bf v} $, we shall find
the generalized momentum $ {\bf p} $ of the particle in
the de Sitter spacetime. Since the Lagrange function
does not depend on the radius vector of the particle $ {\bf r} $,
the generalized momentum is conserved:
\beq
    {\bf p}=\frac{ma{\bf v}}{\sqrt{1-v^2}}=const.
\eeq
Solving the expression (3) for $ {\bf v} $, we  get the time dependence
of the particle velocity:
\beq
    {\bf v}=\pm\frac{{\bf p}}{\sqrt{m^2a^2(\eta)+{\bf p}^2}}.
\eeq
By substituting into (4) the function $ a(\eta) $ and per\-form\-ing
integration, we find  the particle motion law
in the de Sitter space:
\beq
{\bf r}-{\bf r}_{0}=\frac{{\bf p}}{p^2}\biggl{(}\sqrt{m^2
{\alpha}^2+p^2{\eta}^2}-m\alpha\biggr{)}.
\eeq
Calculating the squared 4-vector of the particle ac\-ce\-le\-ra\-tion
(relative to the flat metric), we obtain
\beq
w^{i}w_{i}=-\frac{\dot{\bf v}^2-{\bigl{[}{\bf v},\dot{\bf v}\bigr{]}
}^2}{\bigl{(}1-v^2\bigr{)}}=-\frac{p^2}{m^2{\alpha}^2}=-w^2
\eeq
where
$$
 {\bf w}=\frac {{\bf p}}{m\alpha} = const
$$
is the particle acceleration vector in the intrinsic frame of reference.
It follows from (6) that the particle motion in the de Sitter universe,
considered relative to the conformally related
Minkowski metric, is the relativistic uniformly accelerated motion [2].

In the massless limit Eq. (5) takes the form
\beq
{\bf r}={\bf r}_{0}+{\bf n}|\eta|,\ \ \ \ \  {\bf n}={\bf p}/{p}.
\eeq
Thus, generally speaking, in the massless limit the world line of the
particle differs from a null geodesic since the direction of the particle
velocity vector at the moment of time $ \eta = 0 $ instantly
changes into the opposite.

\section{The field of a charged particle}

Now we will find the electric and magnetic fields created
by a charge moving in the de Sitter universe. For this purpose we consider
the Maxwell equations in a Riemannian spacetime \cite{2}
\beq
F_{ik}^{\ \ ,k}=J_{i},\ \ \ \ \  F_{ik}=A_{i:k}-A_{k:i}
\eeq
with a nonzero right-hand side.
Here $ J^i$ is the 4-vector of current density,
\beq
J^{i}=\frac{e}{\sqrt{-g}}\delta
\bigl{(}{\bf r}-{\bf r}_{e}\bigr{)}\frac{\partial x^{i}}{\partial \eta},
\eeq
${\bf r}_{e}$ is the radius vector of the charge $ e $ and
the electromagnetic
field potential $ A_ {i}$ satisfies the Lorentz condition
\beq
    A_{i:}^{\ \ i}=0
\eeq
(in Eqs. (8) and further rational units of charge are used).
\par
Let us perform in (8)--(10) the conformal trans\-for\-ma\-ti\-on
$g_{ik}=a^2(\eta)\eta_{ik}$, where $\eta_{ik}$ is
the Minkowski metric. The Maxwell equations are invariant under
this transformation, and the Lorentz condition, as a result of
the transformation, takes the form
\beq
    \eta^{ik}\frac{\partial A_{k}}{\partial x^{i}}=-2HA_{0}.       
\eeq
Here $ H = \dot {a} /a $ is the Hubble parameter and
the dot denotes time differentiation.  Substituting $F_ {ik} $
expressed in terms of the potentials into the transformed Maxwell equations
and taking into account the Lorentz condition (11), we  obtain the
equations determining the electromagnetic field potential in the
Robertson-Walker spacetime
\beq
    \eta^{ij}\frac{{\partial}^2 A_{k}}{\partial x^{i}\partial
        x^{j}}+2\frac{\partial}{\partial x^{k}}\Bigl{(}HA_{0}\Bigr{)}=j_{k}
\eeq
where
\beq
j_{k}=\eta_{kl}\rho\frac{dx^{l}}{d\eta}, \ \ \ \ \  \rho=e\delta(\bf{r}-\bf{r}_{e}).
\eeq

Consider the solution of Eqs. (12) in the de Sitter universe. Putting
\beq
    A_{0}=\eta B_{0},
\eeq
we find that the function $ B_0({\bf r}, \eta) $ is a solution of
the inhomogeneous wave equation
$$
    \frac{{\partial}^{2}B_{0}}{\partial{\eta}^2}-\bigtriangleup B_{0}=
        \frac{j_{0}}{\eta}
$$
and hence can be written down as \cite{3}
\beq                                   
B_{0}({\bf r},\eta)=\frac{1}{4\pi}\int\frac{d{\eta}'
           d^{3}x'j_{0}({\bf r}',{\eta}')\delta
            \bigl{(}{\eta}'-\eta+|{\bf r}-{\bf r}'|
            \bigr{)}}{{\eta}'|{\bf r}-{\bf r}'|}. 
\eeq
Having calculated, with the help of (14)--(15),
the scalar potential $A_0$ of the field of the point
charge, we present it as
\beq
A_{0}({\bf r},\eta)=A_{0}^{{}^M}({\bf r},\eta)+A_{0}^{{}^S}({\bf r},\eta).
\eeq
Here $ A_{0}^{{}^M}({\bf r}, \eta) $ is the scalar potential
of the field cre\-a\-ted by the charge
in accelerated motion in flat spacetime
(the Li\'enart-Wiechert potentials) \cite{2}
\[ \begin{array}{crc}
A_{0}^{{}^M}({\bf r},\eta)=\frac{{e}}{4\pi(R-{\bf Rv})},\\
R  =|{\bf r}-{\bf r}_{e}(\eta')|,\\
    {\eta}'+ |{\bf r}-{\bf r}_{e}({\eta}')|=\eta, &
    {\bf v}=\dot{\bf r}_{e}({\eta}').
\end{array} \]
The additional term $ A_{0}^{{}^S} ({\bf r}, \eta)$ is caused
by the space\-time curvature:
$$
    A_{0}^{{}^S}=\frac{R}{{\eta}'(R-{\bf Rv})}.
$$
Now consider the equation for the vector po\-ten\-tial~${\bf A}$
$$
\frac{{\partial}^{2}A_{\alpha}}{\partial{\eta}^{2}}-\bigtriangleup
A_{\alpha}-2\frac{\partial B_{0}}{\partial x^{\alpha}}=j_{\alpha} \ \ \ \ \ 
            (\alpha=1,2,3).
$$
Its solution can be written down as
\beq
    A_{\alpha}=\frac{\partial\varphi}{\partial x^{\alpha}}+A_{\alpha}^{{}^M}  
\eeq
where
${\bf{A}}^{{}^M}=\bigl{(}-A_{1}^{{}^M},-A_{2}^{{}^M},-A_{3}^{{}^M}\bigr{)}$
is the Li\'enart-Wiechert vector potential
$$
    {\bf A}^{{}^M}=\frac{e}{4\pi}\frac{{\bf v}}{(R-{\bf Rv})},
$$
and the function $\varphi({\bf r},\eta)$  has the form
$$
    \varphi=\frac{e}{8{\pi}^2}\int\frac{d^{3}x'}{|{\bf r}-
    {\bf r}'|\eta''(R'- {\bf R}'{\bf v}')}.
$$
Here
$$
{\bf R}'={\bf r}'-{\bf r}_{e}(\eta''),\ \ \ \ \  {\bf v}'= \dot{\bf r}_e(\eta''),
$$
and the moment of time $ {\eta}''$ is determined by the equation
$$
{\eta}''-\eta+|\bf{r}-{\bf r}'| +|{\bf r}'-{\bf r}_{e}({\eta}'')|=0.
$$
Having calculated the electromagnetic field potentials, we can find
the electric and magnetic field strengths using the formulae
\beq
{\bf H}=rot{\bf A}=rot\Bigl{(}{\bf A}^{{}^M}-grad\varphi\Bigr{)}=
                rot{\bf A}^{{}^M}={\bf H}^{{}^M},                 
\eeq
\beq
{\bf E}=-\frac{\partial{\bf A}}{\partial\eta}-grad A_{0}.                     
\eeq
It follows from (18) that the magnetic field strength $ {\bf H} $
created by the point charge moving in the  de
Sitter universe has precisely
the same form as the strength $ {\bf H}^{{}^M} $ of the field
of the charge moving in the Minkowski spacetime. In turn, Eq. (19)
for the electric field strength $ {\bf E} $ can be presented  as
\beq
    {\bf E}={\bf E}^{{}^M}+{\bf E}^{{}^S},     
\eeq
where $ {\bf E}^{{}^M} $ is the
field strength of the point charge moving in
flat spacetime, and the additional term $ {\bf E}^{{}^S} $ has the form
$$
{\bf E}^{{}^S}=grad\biggl(\frac{\partial\varphi}{\partial\eta}- A_{0}^{{}^S}\biggr).
$$
Let us consider the field $ {\bf E}^{{}^S} $ at large distances from the
charge. Calculations show that for $ r\gg r' $,
$ R\gg r_{e} (\eta') $
the field $ {\bf E}^{{}^S} $ is oriented along
the radius vector $ {\bf r} $ of the observation point:
$$
    {\bf E}^{{}^S} = \frac {{\bf r}} {r^2 (\eta-r)} E (\eta-r)
$$
and hence does not give any contribution to the total intensity of
charge radiation.

Thus the radiation field  created by a free charge
moving in the de  Sitter universe is described by the same
formulae as the radiation field of a charge moving with
acceleration in Minkowski spacetime.

\section{The spectrum of radiation}

Now we find the spectrum of radiation of a charge moving  in the de
Sitter universe. The amount of energy $ dI $ propagating per unit of time
through the element $ df = a ^ {2} (\eta) r^{2} d\Omega $ of a spherical
surface centred at the origin is determined by the relation
$$
    dI=T^{0\alpha}df_{\alpha} \ \ \ \ \  (\alpha=1,2,3)
$$
where $ T^{ik} $ is the energy-momentum tensor
of the electromagnetic field
$$
    T^{ik}=-F^{i}_{\ m}F^{km}+\frac{1}{4}g^{ik}F_{mn}F^{mn}.
$$
Expressing its components $ T^{0\alpha} $ in terms of the
electric and magnetic field strengths, we find that the radiation intensity
in the element of solid angle $ d\Omega $ in the de  Sitter universe has the
form
\beq
    dI=\Bigl([{\bf EH}]{\bf n}\Bigr) r^{2}d\Omega.
\eeq
As shown above, in a wave zone the field strengths ${\bf E}$ and $ {\bf H}$
coincide with those produced by an accelerated charge
in flat spacetime. Hence, when calculating the spectral distribution of
radiation energy from a charge in the de  Sitter spacetime, we can use the
result obtained in the framework of classical electrodynamics in
flat spacetime \cite{2}. Namely, the particle radiation energy spectrum
is determined by the expression
\beq
    dE_{\bf k}=\frac{1}{16{\pi}^{3}}j_{m}(k)j^{m\ast}(k)d^{3}k
\eeq
where $ j ^ {m} (k) $ is the Fourier component
of the 4-vector of current density
\beq
    j^{k}=e\int dx^{k}(\eta)
    \exp{\Bigl{(}ik_{0}\eta-i{\bf kr}(\eta)\Bigr{)}}  
\eeq
In (23) the integration is carried out along the particle world line.
If one takes into account that the particle motion law
has the form (5), then the calculation
of the particle radiation spectrum in  the de Sitter universe
will be reduced to the calculation of radiation spectrum of a uniformly
accelerated charge in Minkowski spacetime. The solution of this problem was
obtained by A.I. Nikishov and V.I. Ritus as a special case  when considering
a more general problem of radiation spectrum of a classical
charged particle moving in a constant electric field \cite{4}.
Expressing in Nikishov's formula the field strength $ {\bf E} $ in terms of
the acceleration $w$ and putting $w = p / (m\alpha)$, we find the spectral
distribution of electron radiation energy in the de Sitter spacetime:
\beq
    dE_{\bf k}=\frac{e^{2}}{4{\pi}^3w^2}{\cal K}_{1}^{2}
        \biggl( \frac{k_{\perp}}{w} \biggr) d^3k.
\eeq
Here $ k _ {\perp} $ is the wave vector component perpendi\-cu\-lar
to the particle momentum  $ {\bf p}$ and ${\cal K}_1(a)$ is
the Macdonald function \cite{5}.

Let us now investigate the de\-pen\-dence of the spec\-tral
dis\-tri\-bu\-tion (24) on the spacetime curvature.
If the curvature is small, $ R = 12 {\alpha} ^ {-2} \ll m ^ 2 $,
then $ w\to 0 $ and the radiation spectrum takes the form
\beq
dE_{\bf k}=\frac{e^2dk_{\perp}dk_{\parallel}d\varphi}{8{\pi}^{2}w}
\exp{\Biggr{(}-\frac{2k_{\perp}}{w}\Biggl{)}}
\eeq
where $ k _ {\parallel} $ is the projection of the wave vector on the
direction of the momentum $ {\bf p} $.  As it was to be ex\-pec\-ted, $ dE
_{\bf k} \to 0 $ when $R\to 0$.

In the opposite limiting case, when $ m\alpha\to 0 $, we  obtain
\beq
    dE_{\bf k}=\frac{e^{2}d^{3}k}{4{\pi}^{3}k_{\perp}^{2}}.   
\eeq
It follows from (26) that the radiation spectral density
does not depend on the curvature $ R $ of the de  Sitter universe when
$ R\to \infty $, $ k\ll R $.  Eq. (26) remains valid
for arbitrary values of $ R\neq 0 $ if  $ m\rightarrow 0 $.
Thus, in the massless limit the radiation spectral density
does not tend to zero. The behaviour of the spectral density
in the massless limit is explained by the particle world line being
different from a null geodesic when $ m\to 0 $, the particle experiencing an
instantaneous acceleration at the moment $ \eta = 0 $.

Having  executed, with the help of Eq. (2.16.33.2) from \cite{6},
the integration in
(24) in $k_{0} $, we find the angular distribution of radiation of
an electron in the de Sitter spacetime. The result has the form
\beq
    dE=\frac{3e^2}{128\pi}\frac{wd\Omega}{\sin^{3}\theta},
\eeq
i.e. the energy emitted by the
particle into the solid angle $ d\Omega $ for the whole time
of the motion is proportional to an intrinsic acceleration of the particle
relative to the  Minkowski metric.

It follows from (27) that the total ra\-di\-a\-tion energy for infinite
time from the whole particle trajectory is
infinite. The divergence of the total radiation energy for infinite
time of motion is easy to be un\-der\-stood by taking into account that
according to the Larmor formula the source radiation intensity is finite.

\section{Conclusion}

We have carried out the calculation of radiation spectrum of a classical
charged particle in de Sitter spacetime. It seems to be
of interest to compare Eq. (24) with the results of quantum-mechanical
calculations of the electron radiation spectrum in the spacetime with
the metric (1)--(2).

The results of quantum mechanical calculations obviously
depend on the choice of the vacuum state of the quantum field. If,
for example, we use the Bunch-Davies vacuum [1]  as
the initial and the final vacuum of the spinor field, then we
obtain in the classical limit ($w\to 0 $,
$ k _ {0} \ll p $, $ k _ {0} /w = const$) the
spectral distribution of radiation energy to be \cite{7}
\beq
 dE_{\bf k}=\frac{e^2}{16\pi w^2}{\biggl{[}{\bf
 L}_{-1}\biggl{(}\frac{k_{\perp}}
 {w}\biggl{)}-I_{1}\biggl{(}\frac{k_{\perp}}{w}\biggr{)}\biggr{]}}^2 d^3 k
\eeq
where $I_1(z) $ is the modified Bessel function of the first
kind and $ {\bf L}_{-1} (z) $ is the modified Struve function  [6]. As we
see, the classical limit of the electron radiation spectrum (28) does
not coincide with (24). In our point of view, it means thet the
vacuum state definition in the de Sitter  universe (1)--(2) requires a
specification. However, in the present paper we would like to be restricted
to ascertaining the fact that the use of the Bunch-Davies vacuum for
calculations of quantum effects in the de  Sitter universe leads to the
results that, in the classical limit, contradict the results of the
classical theory. A discussion of the  problem of vacuum choice
in the spacetime  (1)--(2) exceeds the limits of our article. This problem
and also the results of calculations of the electron radiation spectrum in
the de Sitter universe in the framework of quantum electrodynamics in an
external gravitational field will be considered in our next paper.

\subsection*{Acknowledgment}
{The authors would like to express their thanks to V.G. Bagrov and I.L.
Buchbinder for helpful dis\-cus\-sions.}

\end{document}